\documentclass[a4paper,preprint,preprintnumbers, superscriptaddress,showpacs,showkeys,nofootinbib,10pt,twocolumn]{revtex4-1}
\usepackage[english]{babel}
\usepackage[dvips]{graphicx}
\usepackage[breaklinks=true,colorlinks,citecolor=blue,linkcolor=blue,urlcolor=blue]{hyperref}
\usepackage{xcolor,amsthm,amssymb,amsmath,mathtools,latexsym,appendix,physics,bbm}
\usepackage[utf8]{inputenc}

\begin{document}

\title[MSF vs MAF as proper entanglement measures]{Maximal singlet fraction vs. maximal achievable fidelity as proper entanglement measures}
\author{Hermann L. Albrecht Q.}
\email{albrecht@usb.ve}
\affiliation{Departamento de F\'{\i}sica, Universidad Sim\'on Bol\'{\i}var, AP 89000, Caracas 1080, Venezuela.}

\author{Douglas F. Mundarain}
\email[Corresponding Author: ]{dmundarain@ucn.cl}
\affiliation{Departamento de F\'{\i}sica, Universidad Cat\'olica del Norte, Casilla, 1280 Antofagasta, Chile.}

\begin{abstract}
We review some parametric families of 2-qubit states for which concurrence and maximal singlet fraction (MSF) have different and even opposite behaviour. For states considered in this work, maximal achievable fidelity (MFA), a quantity derived by Verstraete {\it et al.} in \cite{Verstraete-MAF}, shows a better agreement with concurrence and complies with important features required to be considered a good entanglement measure.
\end{abstract}
\preprint{SB/F/488-19}
\maketitle


\section{Introduction}
Entanglement is at the very heart of quantum mechanics and is not ``one but rather \emph{the} characteristic trait of quantum mechanics, the one that enforces its entire departure from classical lines of thought'' \cite{schrodinger_1935}, \cite{schrodinger_1936} \cite{Horodecki-Ent}. Many applications of quantum communication and quantum computation use entanglement as its primordial resource for performing certain tasks that classical resources cannot do as well \cite{Bennett_Teleport}, \cite{BB84}, \cite{Ekert_Crypto}. Although it has been shown that non entangled states may display non locality and can be used in quantum information theory \cite{qDiscord-Olliver}, \cite{qDiscord-Henderson}, \cite{Modi-qDiscord}, \cite{qNolocal_Ent}, \cite{ExpQC_NoEnt} entanglement remains of the utmost importance in the field. The study of entanglement and its measures remains a highly active research field and has great relevance to understanding the important role played by quantum mechanics in allowing a more complete manipulation of information. Two of the most widely used entanglement measures are Concurrence \cite{Wooters_Concurrence} and Entanglement Fidelity (or Maximal Singlet Fraction, MSF) \cite{Jozsa_Fidelity}, \cite{Uhlmann_Fid}.

In this work we review the behaviour of both Concurrence and MSF in some particular state families and show that in some cases they can display opposite behaviour. This implies that a particular state appears to be more entangled than another, depending on the use of concurrence or MSF as an entanglement measure. Since it has been well established that concurrence is indeed a good entanglement measure, MSF has to be corrected in order to comply with concurrence.

\section{Maximal Singlet Fraction as an entanglement measure}
In \cite{xiang} a comparison between concurrence and entanglement fidelity is presented for a system composed of two entangled qubits, with the usual AB notation for its subsystems, and a third control qubit, denoted by C, which interacts only with qubit A via the following Hamiltonian:
\begin{equation}\label{eq:HamiltonianoC}
 H = \frac{\lambda}{2}\, \sigma_z^{A}\otimes\vb{I}_2^B\otimes\qty(\dyad{\alpha}-\dyad{\beta})
\end{equation}
\noindent{}where $\lambda $ is a parameter that describes the interaction's strength, $\sigma_z$ is the well known Pauli matrix, $\vb{I}_2^B$ is the identity operator acting o the density matrix ofn B and $\ket{\alpha}$ and $\ket{\beta}$ are two orthonormal states of C. A particular case studied in \cite{xiang} assumes the following initial state for the 3-qubit system:
\begin{equation}
 W = \rho^{AB}_- \otimes \frac{1}{2}\, \vb{I}_2^{C}\,,
\end{equation}
\noindent{}where  $\rho^{AB}_-$ is the density matrix associated  with the singlet $\ket{\Psi^-}=\frac{1}{\sqrt{2}}\,\qty(\ket{01}-\ket{10})$. After interacting via (\ref{eq:HamiltonianoC}), the  reduced density matrix of the AB system at time $t$  has the following structure:
\begin{equation}\label{evolvedstate}
 \rho_{AB} (t) = \cos^2 \left(\frac{\lambda t}{2}\right) \rho^{AB}_-  +\sin^2 \left(\frac{\lambda t}{2}\right) \rho^{AB}_+\,,
\end{equation}
\noindent{}where  $ \rho^{AB}_+$ is the density matrix of one of the Bell states in the triplet $\ket{\Psi^+}=\frac{1}{\sqrt{2}}\,\qty(\ket{01}+\ket{10})$.

\begin{figure}[h]
 \includegraphics[scale=0.6]{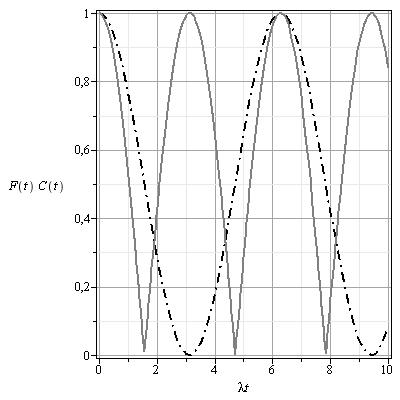}
\caption{\label{fig1}Concurrence (solid line) and Fidelity (dashed line) for state (\ref{evolvedstate}). Graph equivalent to the one presented in  \cite{xiang}.}
\end{figure}

The entanglement fidelity, i.e. MSF, of this state, defined as
\begin{equation}
    F(\rho)\equiv\expval{\rho}{\Psi^-}\,,
\end{equation}
is given by:
\begin{equation}
 F(t)= \cos^2 \left(\frac{\lambda t}{2}\right)
\end{equation}
\noindent{}while its concurrence is given by
\begin{equation}
 C(t) = \left|\cos \left( \lambda t \right)\right|
\end{equation}
In figure (\ref{fig1}) the behaviour of both quantities is graphically presented and their different and even contradictory values can easily be observed.

Due to this results, the authors were compelled to consider a modified version of entanglement fidelity  that incorporates  an optimization process using  local unitary transformations on A and B  respectively. It can be shown that this modified version of entanglement fidelity has analogous  behaviour than  concurrence. The particular modification applied together with the initial conditions considered, implies that the modified  version of entanglement fidelity
becomes a well known quantity: the fully entangled fraction (FEF),
\begin{equation}
    f(\rho)\equiv\max_{\ket{\Psi}}\expval{\rho}{\Psi}\,,
\end{equation}
where $\ket{\Psi}$ in one of the four well known Bell states $\qty{\ket{\psi^\pm},\ket{\phi^\pm}}$.

Using the Bloch-Fano representation, a general formula that allows calculating this quantity was derived  by Badziag {\it et al.} in \cite{badziag}. For any given 2-qubit system with density matrix $\rho$, its Bloch representation is:

\begin{equation}\label{eq:BlochRep}
    \rho = \frac{1}{4} \Bigg( \vb{I}_4 + \va{r}\cdot \va*{\sigma} \otimes \vb{I}_2^{B} +\vb{I}_2^{A} \otimes \va{s}\cdot \va*{\sigma} + \sum_{n,m=1}^3 T_{nm}
\sigma_n \otimes \sigma_m  \Bigg)
\end{equation}

Since local unitary transformations do not modify a system's entanglement, an equivalent state $\rho'$ with a $T'$ diagonal quantum correlation matrix can be obtained:

\begin{equation}\label{eq:BlochRep-TDiagonal}
 \rho'  = \frac{1}{4} \Bigg(\vb{I}_4 + \va{r}\,'\cdot \va*{\sigma} \otimes \vb{I}_2^{B} +\vb{I}_2^{A} \otimes \va{s}\,'\cdot \va*{\sigma}+ \sum_{n=1}^3 \, T'_{n} \,
\sigma_n \otimes \sigma_n \Bigg)
\end{equation}

The fully entangled fraction (FEF) can be evaluated in terms of this quantum correlation matrix as follows:

\begin{equation}
 f = \left\{ \begin{array}{cc}
          \frac{1}{4} \left( 1+ \sum_i |T'_i| \right)&\rm{if}\;  \det(\vb{T}') \leq 0\\
\\
\frac{1}{4} \left[ 1+ {\max}\left(|T'_i|+|T'_j|-|T'_k|\right)\right]&\rm{if}\;  \det(\vb{T}') >0\end{array}
\right.
\end{equation}

State \eqref{evolvedstate} is a Bell Diagonal state with diagonal quantum correlation matrix:

\begin{equation}
 \vb{T} = \mqty(\dmat[0]{
2 \cos^2 \qty(\frac{\lambda t}{2}) -1, 2 \cos^2 \qty(\frac{\lambda t}{2}) -1,-1})
\end{equation}

\noindent{}Since $\det(\vb{T})$ is strictly less than zero, the FEF is

\begin{equation}\label{eq:f}
 f= \frac{1}{2} \left[ 1+\left|2 \cos^2 \left( \frac{\lambda t}{2}  \right)-1\right| \right]
\end{equation}

\begin{figure}[h]
 \includegraphics[scale=0.6]{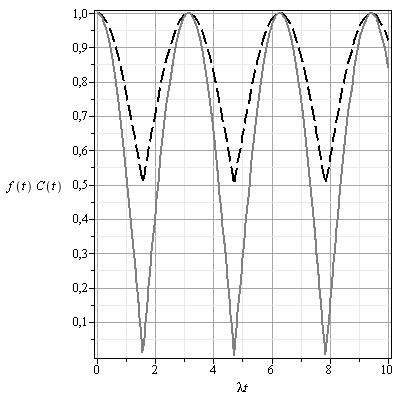}
\caption{\label{fig2}Concurrence (solid line) and MSF (dashed line) for state (\ref{evolvedstate}). Graph equivalent to the one presented in \cite{xiang}.}
\end{figure}
In figure \ref{fig2}) the behaviour of both $f(t)$ and concurrence $C(t)$ is graphically presented. In it, similarities for both quantities can clearly be seen. This result might lead to wrongfully conclude that concurrence and FEF share similar behaviour and therefore it ought to be considered a good entanglement measure. This is certainly not the case. As demonstrated by Verstraete and Verschelde in \cite{Verstraete_Fid}, FEF complies both
\begin{equation} \label{dada2}
 f \geq \max \left(\frac{1+C}{4},C\right)
\end{equation}
\noindent{}as well as
\begin{equation} \label{dada1}
 f \leq \frac{1+C}{2}\,.
\end{equation}
Therefore, for a given value of  $f\in \left(\case{1}{4},\case{1}{2}\right]$, the state can be either entangled or not. Hence, FEF is not an entanglement monotone \cite{EntMono}, i.e. not a good entanglement measure.

\section{Maximal singlet fraction, concurrence and entanglement monotones}

In this section we will present examples of bipartite systems that show opposing behaviour for their MSF and concurrence. First, we consider the example presented in \cite{badziag}, where one of the subsystems interacts locally with the environment, modelled as an Amplitude Damping Channel. Consider the state:
\begin{widetext}
\begin{equation}\label{state6}
\rho = \left(
\begin{array}{cccc}
 p \left( \frac{3}{2} -\sqrt{2}\right) &0&0&0\\
\\
0& \left(1-p \right) \left( \frac{3}{2} -\sqrt{2}\right)&\sqrt{1-p} \left( \frac{1}{2} -\frac{\sqrt{2}}{2} \right)&0\\
\\
0&\sqrt{1-p} \left( \frac{1}{2} -\frac{\sqrt{2}}{2} \right)& \frac{1}{2} +p \left(\sqrt{2}-1\right)&0\\
\\
0&0&0&\left(1-p \right)\left(\sqrt{2}-1\right)
\end{array}
\right)
\end{equation}
\end{widetext}
\noindent{}where $p\in [0,1]$ is the interaction parameter. For $p=0$, $\rho$ is the initial state presented in \cite{badziag}, whose only non-zero Bloch parameters are $x_3=2\qty(1-\sqrt{2})$, $T_{11}=T_{22}=1-\sqrt{2}$ and $T_{33}=2\sqrt{2}-3$. For this state, FEF is
\begin{equation}
f(p)= 1-\frac{\sqrt{2}}{2}+p \left(\sqrt{2}-\frac{5}{4}\right)+\frac{\sqrt{1-p}}{2}\left( \sqrt{2}-1 \right)
\end{equation}

\noindent{}while its  concurrence is given by
\begin{equation}
C(p) = \sqrt{1-p}(\sqrt{2}-1) +(\sqrt{2}-2) \sqrt{\sqrt{2}-1} \sqrt{p (1-p)}
\end{equation}
\begin{figure}[h]
 \includegraphics[scale=0.6]{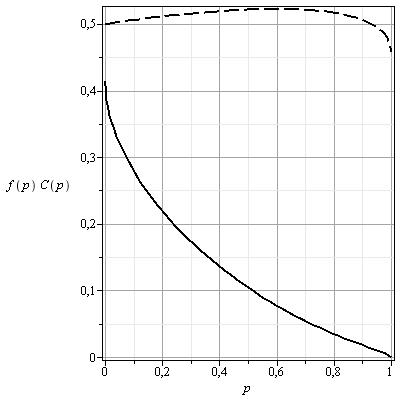}
\caption{\label{fig3}MSF (dashed line) and Concurrence (solid line) for state (\ref{state6}).}
\end{figure}

In figure (\ref{fig3}) both concurrence and FEF are presented as functions of the interaction parameter $p$.  While concurrence diminishes for all $p$, there is an important range of values for which MSF steadily grows, i.e. $p\in \left[0, \left(\sqrt{2}-1\right)/\left(2\sqrt{2}-5\right)\sim 0, 6023\right]$.

\begin{figure}[h]
 \includegraphics[scale=0.6]{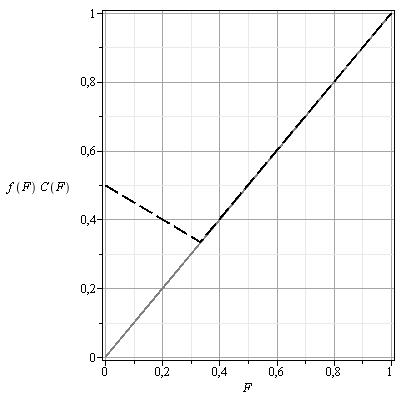}
\caption{\label{fig4}MSF (solid line) and Concurrence (dashed line) for state (\ref{state7}).}
\end{figure}

Consider now the following X state, presented in \cite{Verstraete-MAF}:
\begin{equation}\label{state7}
 \rho = F \dyad{\Phi^+} +(1-F)\dyad{01}
\end{equation}
\noindent{}where $\ket{\Phi^+} =\frac{1}{\sqrt{2}} \qty(\ket{00}+\ket{11})$ is the well known Bell state. Its concurrence is given by
\begin{equation}
 C = F
\end{equation}
\noindent{}while its FEF is
\begin{equation}
 f = \left\{\begin{array}{cc}
          \frac{1-F}{2} & {\rm if}\qquad 0 \leq F \leq \frac{1}{3}\\
                      \\
        F & {\rm if} \qquad  \frac{1}{3} \leq F \leq 1
            \end{array}
\right.
\end{equation}
Again, for  $0 \leq F \leq \case{1}{3}$, there is an opposing behaviour for the systems concurrence, which steadily decreases, and its FEF, which in turn grows, as shown in figure (\ref{fig4}). Therefore, in order to develop a good entanglement measure related to FEF, new concepts need to be introduced.

\section{Maximal Achievable Fidelity}
Such a quantity related to FEF that shows similar behaviour than concurrence is the maximal fidelity achievable by trace-preserving LOCC operations or simply maximal achievable fidelity (MAF) \cite{Verstraete-MAF}, which satisfies important features required for \emph{bona fide} entanglement measures, e.g. it is an entanglement monotone. Recently, MAF was studied in the context of noisy quantum channels by Bandyopadhyay et al. \cite{Bandyopadhyay2012}, \cite{Bandyopadhyay2014}. 

This optimal entanglement fidelity is defined by a convex semi definite program:
\begin{eqnarray}
&&\mathrm{Maximize}\nonumber\\
&&f^* = \frac{1}{2} - \mathrm{Tr} \left( X \rho^{\Gamma} \right)\\
&&\mathrm{under\, the\, constrains}\nonumber\\
&&0 \leq X\leq \mathbf{I} \qquad  -\frac{\mathbf{I}}{2} \leq X^{\Gamma} \leq \frac{\mathbf{I}}{2}
\end{eqnarray}
\noindent{}where $X$ is a local state-dependent filter to be applied to one of the subsystems within the optimization LOCC protocol. For state (\ref{state7}), this optimization program gives:

\begin{equation}\label{eq:f*Verstraete}
 f^{*}_{op} = \left\{\begin{array}{cc}
                        \frac{1}{2} \left[ 1+\frac{F^2}{4 ( 1-F)} \right] & {\rm if} \qquad  0\leq F \leq \frac{2}{3}\\
                        \\
                        F & {\rm if}\qquad \frac{2}{3} \leq F \leq 1
                    \end{array}\right.
\end{equation}
\indent{}In \cite{Verstraete-MAF}, results are reported for $1/3\geq F\geq 1$ only. After some analysis, it can be demonstrated that the first expression in \eqref{eq:f*Verstraete} holds also for $0\geq F\geq 1/3$ and therefore $f^{*}_{op}\,\forall\,F\in[0,1]$ is as stated. For completeness, LOCC operations required to obtaining this MAF are shown in the appendix.

In Figure (\ref{fig5}), MAF, concurrence and FEF are presented graphically. Both MAF and concurrence display similar behaviour, i.e. both of them are increasing functions of parameter F.
\begin{figure}[h]
 \includegraphics[scale=0.6]{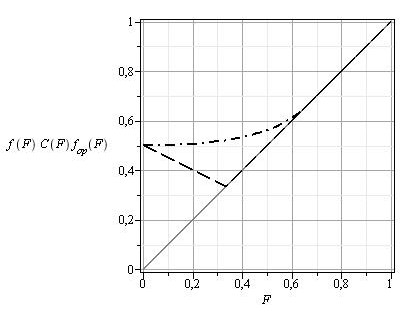}
\caption{\label{fig5}Concurrence (solid line), MSF (dashed line) and MAF (dot-dashed line) for state (\ref{state7}).}
\end{figure}

For state (\ref{state6}) and  from similar calculations, the following results are obtained:
\begin{widetext}
\begin{equation}\label{eq:f*Badziag}
 f^{*}_{op} = \left\{
                    \begin{array}{ll}
                    \frac{-3/2+ \sqrt{2} }{2} p +\frac{\sqrt{2}+3}{8} & {\rm if} \qquad 0\leq p \leq \frac{3}{4}\\
                    \\
                    f &{\rm if} \qquad \frac{3}{4} \leq p \leq \frac{\left(\sqrt{2}+1\right)\left(\sqrt{7+2 \sqrt{2}}-\sqrt{2}-1\right)}{2}\\
                    \\
                    \frac{1+\left(3-2 \sqrt{2}\right)p +2\left(\sqrt{2}-1\right) p^2}{ 4 p}  & {\rm if} \qquad  \frac{\left(\sqrt{2}+1\right)\left(\sqrt{7+2 \sqrt{2}}-\sqrt{2}-1\right)}{2}  \leq p \leq 1\\
                    \end{array}\right.
\end{equation}
\end{widetext}

\begin{figure}[h]
 \includegraphics[scale=0.6]{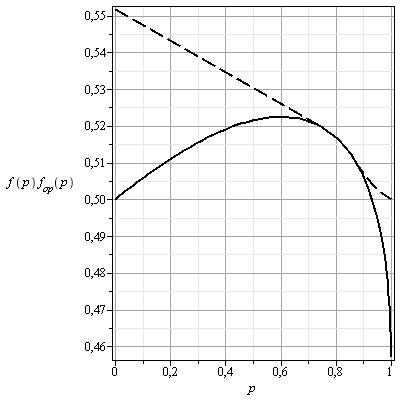}
\caption{ \label{fig6}MSF (solid line) and MAF (dashed line) for state (\ref{state6}).}
\end{figure}

In figure (\ref{fig6}), MSF and MAF for state (\ref{state6}) are shown. Their behaviour is analogous as the one observed in the aforementioned case.

\section{Final remarks: Restrictions}
Some remarks must be made regarding the trace preserving transformation proposed by Verstraete and Verschelde in  \cite{Verstraete-MAF}. Let $C$ be a given concurrence. Then, FEF complies \eqref{dada1}, alongside (\ref{dada2}). For any state satisfying equality in (\ref{dada1}), e.g. Bell Diagonal States, FEF cannot be increased by means of trace preserving transformations. Any increase in FEF in such a case would imply an increase of its concurrence, which is forbidden.

One might be compelled to assume that for states which do not saturate this condition, it is always possible to optimize its FEF. This is nevertheless false. State (\ref{state6}) satisfies the  equality condition only for $p = 2 \sqrt{2}-2 \simeq 0.8284$, and  there is a whole neighbourhood around this value representing a set of states that do not satisfy the equality condition, for which it is not possible to increase FEF, as can be seen in Figure (\ref{fig5}) from the overlapping of FEF and MAF. This is indeed what (\ref{eq:f*Badziag}) implies for $3/4 \leq p \leq \left(\sqrt{2}+1\right)\left(\sqrt{7+2 \sqrt{2}}-\sqrt{2}-1\right)/2$, since in it, $f^{*}_{op}=f$.

\begin{figure}[h]
 \includegraphics[scale=.4]{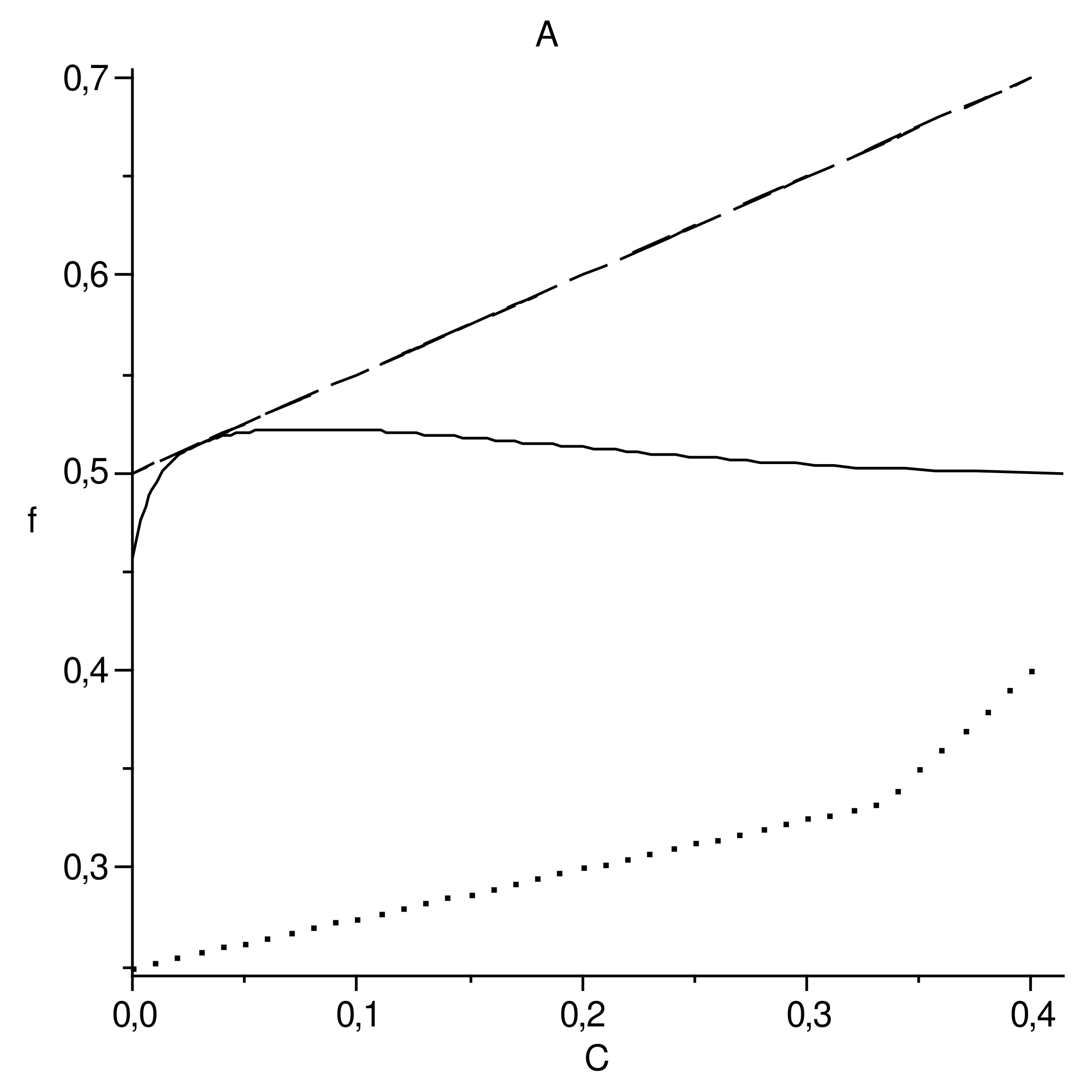}\hspace*{1cm}\\
 \includegraphics[scale=.4]{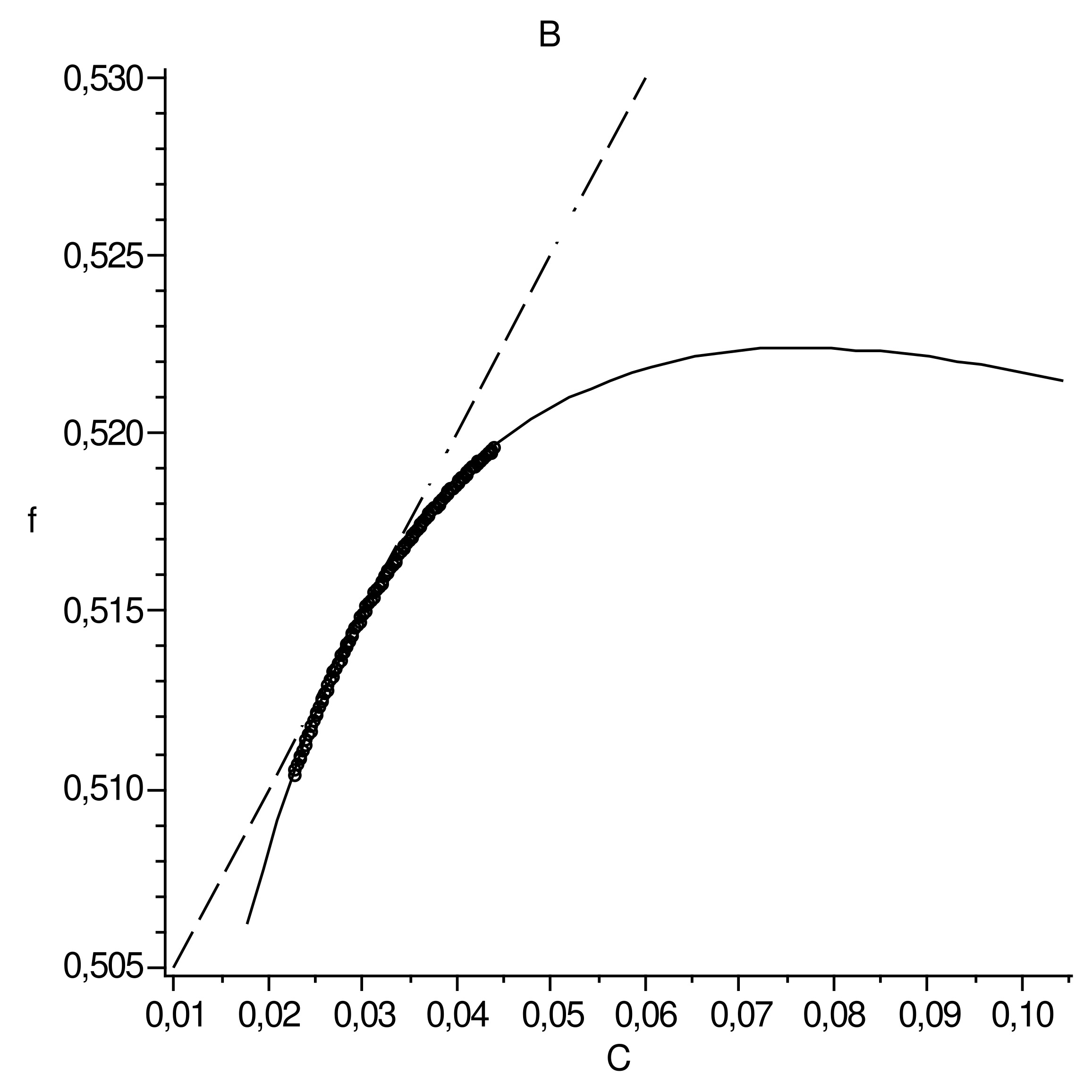}
\caption{MSF vs Concurrence for state (\ref{state6})}
 \label{fig7}
\end{figure}

This effect can also be  observed in Figure (\ref{fig7}) where FEF as a function of concurrence for state (\ref{state6}) is plotted. In it, maximal FEF for a given concurrence is represented by a dashed line and minimal is represented by a point line at the bottom. Figure (\ref{fig7}-B) is a zoom near the state satisfying  equality condition in (\ref{dada1}). The neighbourhood of states whose FEF cannot be optimized are represented by a thicker line.

An additional upper bound to \eqref{dada1} for the MAF is given in terms of its Negativity \cite{Verstraete_Fid} for any generic 2-qubit state $\rho$ as
\begin{equation}\label{eq:f*Nega}
    f^*(\rho) \leq \frac{1}{2}\qty[1+N(\rho)]\,,
\end{equation}
\noindent{}which follows directly from the equivalent bound found for fidelity \cite{Verstraete_Fid}. For state \eqref{state7}, its Negativity is given by

\begin{equation}\label{eq:NegatividadVV}
    N=\max\qty{0,\,F-1+\sqrt{F^2+(F-1)^2}}\,.
\end{equation}

\begin{figure}[h]
\centering
 \includegraphics[width=0.35\textwidth]{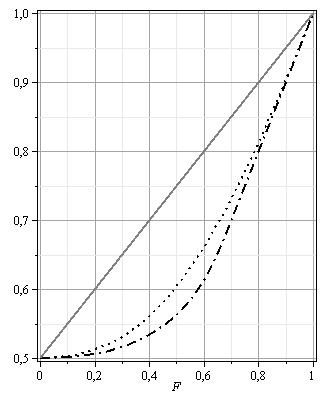}
\caption{\label{fig:MAFvsSup}MAF (dot-dashed line) and upper bounds \eqref{eq:f*Nega}, dotted, and \eqref{dada1}, solid, for state (\ref{state7}).}
\end{figure}

In Figure \ref{fig:MAFvsSup}, MAF along with the upper bounds given in \eqref{dada1} and \eqref{eq:f*Nega}, are presented.

From all this previous analysis, it is derived the fact that there are states that even though for a given concurrence do not reach maximal FEF values, cannot be optimized. This can be observed in a perhaps more straightforward way when state (\ref{state7}) is considered. In this case,  $2/3 \leq  F \leq 1 $ represents a set of states that have minimal FEF values for a given concurrence but cannot be optimized.

The FEF optimization problem is a very complex one. To our knowledge, there is no analytical formula for the increase of the singlet fraction, neither a simple way to {\it{a priori}} determine whether a state can be optimized or not. Only when FEF is maximal for a given concurrence, it can be stated that that set of states cannot be optimized. Moreover, for any set of states with a given concurrence and FEF, such that the strict inequality holds in (\ref{dada1}), there will be subsets of states that can be optimized and subsets for which this is not possible.

MAF, being an entanglement monotone, complies with the main characteristics required to be a good entanglement measure and is therefore a better choice for characterizing a systems entanglement.

\section{Conclusions}

We have reviewed certain families of states whose concurrence and MSF have different and even opposite behaviours. This, along the fact that MSF is not an entanglement monotone, leads to the conclusion that MSF is not a good entanglement measure. On the other hand, MAF is indeed an entanglement monotone, shows analogous behaviour than concurrence and therefore complies with the main characteristics of a good entanglement measure.

\section{Acknowledgments}
Albrecht would like to thank the support given by the research group GID-30, \emph{Teoría de Campos y Óptica Cuántica}, at the Universidad Simón Bolívar, Venezuela.

\bibliographystyle{unsrt}
\bibliography{biblio}
\include{biblio}

\appendix
\appendixpage
Consider the following state  (\ref{state7}):

\begin{equation}
 \rho = F \ket{\Phi^+}\bra{\Phi^+}+(1-F)\ket{01}\bra{01}
\end{equation}

This state was studied in  \cite{Verstraete-MAF}, in which the authors found its MAF for $1/3 \leq F \leq 1$. It is straightforward to verify the extension
for $0 \leq F \leq 1/3$.

\begin{equation}
 f^{*}_{op} = \left\{\begin{array}{cc}
        \frac{1}{2} \left[ 1+\frac{F^2}{4 ( 1-F)} \right] & {\rm if} \qquad  0\leq F \leq 2/3\\
        \\
        F & {\rm if}\qquad 2/3 \leq F \leq 1
\end{array}\right.
\end{equation}
The optimal filtering operation \cite{Verstraete-MAF}, which is a local operation, performed only on subsystem B, has the following structure:
\begin{equation}
\mathbf{A}_1 =\vb{I}_2^{A} \otimes\left[ \ket{0}\bra{0}+\frac{F}{ 2 (1-F)}\ket{1}\bra{1}\right]
\end{equation}
This filtering operation has the following success probability:
\begin{equation}
P_1 =  \frac{F \left(3 F^2-6 F+4\right)}{8 (1-F)^2}
\end{equation}

If the filtering operation is successful, the system state after it is given by
\begin{equation}
 \rho_1 = \frac{1}{P_1} \mathbf{A}_1  \rho  \mathbf{A}^{\dagger}_1
\end{equation}
On the other hand, if the filtering operation is not successful, the state after it is then given by
\begin{equation}
 \rho_2 = \frac{1}{P_2} \mathbf{A}_2 \rho  \mathbf{A}^{\dagger}_2
\end{equation}
where
\begin{equation}
\mathbf{A}_2 = \vb{I}_2^{A}\otimes\left[\sqrt{1-\frac{F^2}{4 (1-F)^2}}\,\ket{0}\bra{1}\right]
\end{equation}
and $P_2=1-P_1$ is the probability of failure, which is given by
\begin{equation}
P_2 =  \frac{(2-F) (3 F^2-8 F+4)}{8 (F-1)^2}
\end{equation}

It is easy to verify that state $\rho_2$ is a separable one, i.e. has null concurrence, but its MSF is different from $1/2$, which is a necessary
condition in order to obtain the MAF. In this case, state
\begin{equation}
 \rho'= \mathbf{A}_1 \rho \mathbf{A}_1^{\dagger}+\mathbf{A}_2 \rho \mathbf{A}_2^{\dagger}
\end{equation}
has no MAF.

In order to get the MAF,  state $\rho_2$,  which is the output state  when the filtering operation is not successful,  must be transformed into a state with MSF equal to $1/2$. This transformation can be performed by LOCC on subsystem A: an observer measures the $\sigma_z$ observable on subsystem B. If eigenvalue  1 is obtained, the system is left unaffected. If the result obtained is 0, the state is to be transformed into one with eigenvalue 1, i.e. state $|1\rangle$. In both cases, the final state of subsystem B is $|1\rangle$. For the whole system, the final state, incorporating success and failure, is therefore given by

\begin{equation}
 \rho''= \mathbf{A}_1 \rho \mathbf{A}_1^{\dagger}+(1-P_1) \rho_0
\end{equation}
with
\begin{equation}
\rho_0 = \ket{00}\bra{00}
\end{equation}

State $\rho''$ has a MSF equal to the MAF and it is straightforward to verify that this state has less concurrence than the initial state $\rho$.
\end{document}